\documentclass[12pt,preprint,usenatbib,eqsecnum]{aastex}

\usepackage{graphicx}
\usepackage{dcolumn}
\usepackage{bm}
\usepackage[]{natbib}
\begin{document}


\title{Perturbative Analysis of Universality and Individuality\\ in Gravitational Waves from Neutron Stars}

\author{L.K. Tsui and P.T. Leung\footnote{Email:
ptleung@phy.cuhk.edu.hk} }
\affil{%
Physics Department and Institute of Theoretical Physics,\\ The
Chinese University of Hong Kong,\\ Shatin, Hong Kong SAR, China.
}%

\date{\today}
\def\tomega{ \tilde{\omega} }
\def\tOmega{ \tilde{\omega} }
\def\tr{ \tilde{r} }
\def\tx{ \tilde{r}_* }
\def\tV{ \tilde{V} }
\def\tpsi{ \tilde{\psi} }
\def\trho{ \tilde{\rho} }
\def\tP{ \tilde{P} }
\def\tR{ \tilde{R} }
\def\tX{ \tilde{R}_* }
\def\tm{ \tilde{m} }
\def\tphi{ \tilde{\nu} }
\def\tlam{ \tilde{\lambda}}
\def\tepsilon { \tilde{\epsilon}}
\def\tU{ \tilde{U} }
\def\tphi{ \tilde{\nu} }
\def\tlam{ \tilde{\lambda}}
\def\tepsilon { \tilde{\epsilon}}
\def\cc{{\cal C}}
\def\a{ {\rm a}}
\def\c{ {\rm c}}
\def\e{ {\rm e}}
\def\p{ {\rm p}}
\def\f{ {\rm f}}
\def\d{ {\rm d}}
\def\i{ {\rm i}}
\def\r{ {\rm r}}
\begin{abstract}
The universality observed in gravitational wave spectra of
non-rotating neutron stars is analyzed here. We show that the
universality in the axial oscillation mode can be reproduced with
a simple stellar model, namely the centrifugal barrier
approximation (CBA), which captures the essence of the Tolman VII
model of compact stars. Through the establishment of scaled
co-ordinate logarithmic perturbation theory (SCLPT), we are able
to explain and quantitatively predict such universal behavior. In
addition, quasi-normal modes of individual neutron stars
characterized by different equations of state can be obtained from
those of CBA with SCLPT.
\end{abstract}


\maketitle
\section{Introduction}
As is well known, neutron stars of high densities are ideal
extra-terrestrial test beds for theories of nuclear matters, quark
matters and high energy physics \citep[see e.g.][and references
therein]{ComStar}. For instance, the possible existence of the
quark star \citep[see e.g.][]{Cheng}, a variant of the neutron
star, could lend direct support to the theory of quark matter
\citep{MITBM,Witten,AFO,PBP}. Besides, the relation between the
mass and the radius of a neutron star could determine, up to
certain accuracy, the equation of state (EOS) of relevant nuclear
matter through the process of inversion \citep{Lindblom_invert}.
Therefore, a comprehensive study of observed data and statistics
of neutron stars is likely to lead to fruitful results in the
fields mentioned above.

Traditional means to gather information about neutron stars have
so far relied on electromagnetic waves emitted from them. However,
with the advent of gravitational wave detectors of various designs
\citep[see e.g.][and references therein]{Hughes_03,Grishchuk},
including resonant antennas (e.g. EXPLOPER and NIBOE),
ground-based interferometers (e.g. LIGO and VIRGO),  and the
space-based interferometer LISA, it is generally believed that
neutron stars, as promising gravitational wave emitters, can be
observed and analyzed in the gravitational wave channel within one
or two decades.  For example, it has been argued that the
frequency of detection of gravitational waves emitted in the
mergers of binary neutron stars  could be as high as several
hundreds per year in the near future
\citep{Belczynski:2001uc,Hughes_03}. Besides, gravitational waves
could also be emitted during the formation of neutron stars
\citep{Lindblom:1998wf,Fryer:2001zw}. Being spurred on by the
possibility of inferring the internal structure of neutron stars
from the gravitational wave signals emitted by them, researchers
have been actively examining the peculiarities embedded in such
signals
\citep{Andersson_1996,Andersson1998,Ferrari,Kokkotas_2001}.

Despite that the evaluation of gravitational waves emitted in
violent stellar activities such as binary mergers and asymmetric
core collapse has indeed posed a grand challenge to researchers in
numerical relativity, linearized theory of pulsating neutron stars
pioneered by \citet{Thorne} can still provide useful insight into
such complex situations. By analogy of radiating electric circuits
\citep{ToyModel}, linearized gravitational waves are analyzed in
terms of quasi-normal modes (QNMs), which are damped harmonic
pulsations with time dependence $\exp(\i\omega t)$ and are
characterized by complex eigenfrequencies
$\omega=\omega_\r+\i\omega_\i$
\citep{Press_1971,Leaver_1986,Ching,Kokkotas_rev,Nollert_rev}. It
is believed that the QNM frequencies of a pulsating neutron star
can reflect the physical characteristics of the star, such as its
mass $M$, radius $R$ and EOS as well
\citep{Andersson_1996,Andersson1998,Ferrari,Kokkotas_2001}.
However, \citet{Andersson1998} and \citet{Ferrari} also noted that
the frequency $\omega_\r$ and the damping time $\tau\equiv
1/\omega_\i$ of the leading axial and polar $w$-modes of
non-rotating neutron stars follow approximately the equations:
\begin{eqnarray}
\omega_\r & \approx & \frac{1}{R}\left[a_\r
\left(\frac{M}{R}\right)+b_\r
\right] ,\label{BBF1}\\
\frac{1}{\tau}& \approx & \frac{1}{M}\left[a_\i
\left(\frac{M}{R}\right)^2+b_\i\left(\frac{M}{R}\right)+c_\i
\right],\label{BBF2}
\end{eqnarray}
where $a_\r$, $b_\r$, $a_\i$, $b_\i$ and $c_\i$ are
model-independent constants determined from curve fitting. These
universal behaviors seem to undermine the possibility of inferring
the characteristics of a neutron star from its $w$-mode QNMs.
Instead, they are indicative of a common feature shared among
neutron stars with different EOS's.

To seek the physical mechanism underlying the universality in the
gravitational wave of neutron stars, we showed in a recent paper
\citep{preprint} that (i) The scaled complex eigenfrequencies
$\tomega \equiv M \omega$ of $w$-mode oscillations, to a good
approximation, depend only on the compactness $\cc \equiv M/R$.
(ii) Eqs.~(\ref{BBF1}) and (\ref{BBF2}) are equivalent to a single
formula of $\tomega$:
\begin{eqnarray}
\tomega \approx a
\left(\frac{M}{R}\right)^2+b\left(\frac{M}{R}\right)+c
,\label{TLA}
\end{eqnarray}
where $a$, $b$ and $c$ are complex constants. (iii) The
universality indicated by (\ref{TLA}) is a direct reflection of
the fact that the the mass distribution inside neutron stars with
different EOS's can be nicely approximated by the Tolman VII model
(TVIIM) proposed by \citet{Tolman:1939jz}.

In this paper we further pinpoint the crux of the mechanism
leading to the universal behavior (\ref{TLA}) for axial $w$-mode
pulsations. Inspired by the success of TVIIM, we propose here yet
another simple approximation, namely the centrifugal barrier
approximation (CBA), in which the potential in the axial
gravitational wave equation \citep{Chandrasekhar1} is replaced by
a standard centrifugal barrier in the tortoise radius plus a
constant determined from continuity of the potential at the
stellar surface. The CBA potential yields the correct asymptotic
behavior of the actual potential at the center of the star and is
a good global approximation as well. The wave function of the
axial gravitational wave equation in CBA then becomes exactly
solvable. It is remarkable that the universal behavior (\ref{TLA})
can still be reproduced with CBA using the tortoise radius of
TVIIM star as an input. Hence, the observed universality in
neutron star axial pulsations is in fact ascribable to (i) the
centrifugal potential at the center of the star; (ii) the
continuity of the potential; and (iii) the tortoise radius of
TVIIM star, which will be given explicitly in Sect.~2 of our
paper.

To understand quantitatively the universal behavior (\ref{TLA}),
we establish here a scaled co-ordinate logarithmic perturbation
theory (SCLPT), which is a generalization of the logarithmic
perturbation theory (LPT) formulated previously to study QNMs of
``dirty black holes" \citep{dirty1,dirty2}. Applying SCLPT to CBA
and using the tortoise radius of TVIIM star, we obtain from first
principle the numerical values of $a$, $b$ and $c$  in
(\ref{TLA}).
 The universal behavior
(\ref{TLA}) is therefore substantiated analytically.

While the CBA star itself clearly displays universality as
mentioned above, we can also  obtain accurate QNM frequencies of
individual realistic neutron stars from those of CBA with SCLPT.
As a consequence, we are able to consider the EOS-dependence of
the scaled QNM frequency $\tomega$ from the static structure of a
star (e.g. mass distribution), thereby paving the way for
inferring the EOS of a star from its gravitational wave spectra.
Thus, the universality and individuality in QNMs of neutron stars
constructed with different EOS's are fully explored with SCLPT
developed in this paper.

The organization of our paper is as follows. In Sect.~2, after
briefly reviewing the scaling behavior of the axial gravitational
wave equation, we introduce CBA and show that under such
approximation QNMs still manifest the generic behavior summarized
in (\ref{TLA}). In Sect.~3 we formulate SCLPT to study how QNM
frequencies are affected by changes in the potential and the
tortoise radius of the star. In Sect.~4 we apply SCLPT to expand
the scaled QNM frequency $\tomega$ as a quadratic function of the
compactness of the star and in turn corroborate the universal
behavior in (\ref{TLA}). In Sect.~5 SCLPT is applied to
investigate the EOS-dependence of the scaled QNM frequency
$\tomega$. We then conclude our paper in Sect.~6 with a discussion
studying the applicability of CBA and SCLPT to $w_{\rm II}$ and
trapped modes \citep[see e.g.][]{Kokkotas_rev,trap1}. Unless
otherwise stated, geometrized units in which $G=c=1$ are adopted
in the present paper and numerical results are shown for the least
damped mode of quadrupole gravitational waves.

\section{Generic behavior of QNMs and CBA}
The eigenvalue equation for QNMs of  axial oscillations of neutron
stars is given by a Regge-Wheeler-type equation
\citep{Chandrasekhar1}:
\begin{equation}
\label{KG_eq} \left[\frac{\d^2}{\d
r_{*}^2}+\omega^2-V_{}(r_{*})\right]\psi_{}(r_{*}) = 0.
\end{equation}
Here the tortoise coordinate $r_{*}$  is related to the
circumferential radius $r$ by
\begin{equation}\label{r*_in}
r_{*}=\int_{0}^r \e^{(-\nu+\lambda)/2} \d r,
\end{equation}
where $e^{\nu(r)}$ and $e^{\lambda(r)}$ are metric coefficients
defined by the line element $\d s$ as follows:
\begin{equation}
\d s^2=-e^{\nu(r)}\d t^2+e^{\lambda(r)} \d r^2+r^2(\d
\theta^2+\sin^2\theta \d \varphi^2).
\end{equation}
In addition, the metric coefficient $e^{-\lambda(r)}$ is given
explicitly  by:
\begin{equation}
e^{-\lambda(r)}=1-\frac{2m(r)}{r},
\end{equation}
with $m(r)$ being the mass-energy inside circumferential radius
$r$.  The potential $V$ inside the star is given by:
\begin{equation}\label{RW_V}
V_{}(r_{*})=\frac{\e^\nu}{r^3}[l(l+1)r+4\pi r^3(\rho-P)-6m(r)],
\end{equation}
where $\rho(r)$ and $P(r)$ are  the mass-energy density and the
pressure at a point $(r,\theta,\varphi)$, respectively. Outside
the star, the tortoise radial coordinate reduces to
\begin{equation}\label{r*_out}
r_{*}=r+2M \ln\Big(\frac{r}{2M}-1\Big)+C,
\end{equation}
where $C$ is a constant that can be obtained by matching
(\ref{r*_in}) with (\ref{r*_out}) at $r=R$, and the potential $V$
is given by the well known Regge-Wheeler potential \citep{RWeq}.
\begin{equation}\label{RWP_out}
V_{\rm
rw}(r_{*})=\left(1-\frac{2M}{r}\right)\left[\frac{l(l+1)}{r^2}-
\frac{6M}{r^3}\right].
\end{equation}

\begin{figure}
\includegraphics[angle=270,width=8.5cm]{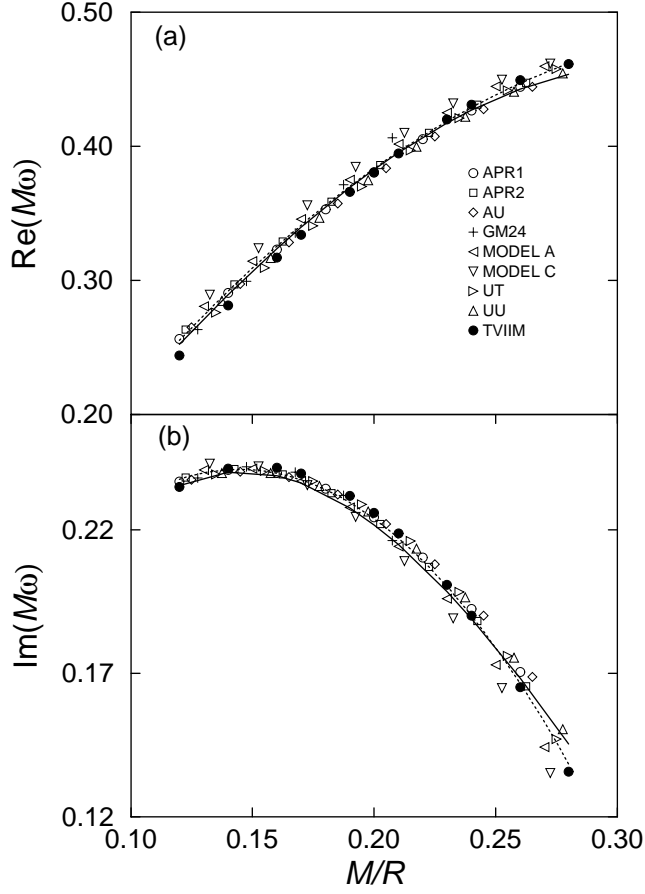}
\caption{The real and imaginary parts of the scaled QNM
frequencies of various realistic stellar models (including APR1,
APR2, AU, GM24, Models A and B, UT and UU), TVIIM and CBA (solid
line) are plotted against the compactness. The dotted line is the
best quadratic fit to those of the realistic stars} \label{f1}
\end{figure}

Eq.~(\ref{KG_eq}), referred to as neutron star Regge-Wheeler
equation (NSRWE) in the following discussion, and the outgoing
wave boundary condition at spatial infinity together determine the
QNM frequency of axial $w$-mode oscillations. We show in
Fig.~\ref{f1} the QNM frequencies obtained numerically for neutron
stars with different EOS's, including models A \citep{modelA} and
C \citep{modelC} proposed by Pandharipande, three models (AU, UU
and UT) proposed by \citet{AU}, models APR1 and APR2 proposed by
\citet{APR}, and model GM${24}$ \citep[][p. 244]{ComStar}. It is
clearly shown that both the real and imaginary parts of the scaled
frequency $\tomega$ are well approximated by quadratic functions
in compactness $\cc$, as indicated by the dotted lines in the
corresponding figures. This in turn leads to the universality
discussed in Sect.~1.

To investigate the physical origin underlying the universality, we
have shown that the axial mode wave equation displays scaling
behavior \citep{preprint}, and can be rewritten as follows:
\begin{equation}
\label{KG_sc}
\left[\frac{\d^2}{\d\tx^2}+\tomega^2-\tV_{}(\tx)\right]\tpsi_{}(\tx)
= 0,
\end{equation}
where
\begin{equation}\label{SP}
\tV_{}(\tx)\equiv M^2 V_{}(\tx)=\left\{%
\begin{array}{ll}
    {\e^{\tphi}}\tr^{-3}\left[l(l+1)\tr+4\pi
\tr^3(\trho-\tP)-6\tm(\tr)\right], & \hbox{$\tr \leq R/M$;} \\
    \left(1-{2}{\tr}^{-1}\right)\left[{l(l+1)}{\tr^{-2}}-
{6}{\tr^{-3}}\right], & \hbox{$\tr > R/M$;} \\
\end{array}%
\right.
\end{equation}
and
\begin{eqnarray}
\tr &=& \frac{r}{M},  \\
\tx &=& \frac{r_*}{M},  \\
\tm(\tr) &=& \frac{m(r)}{M},  \\
\tP(\tr) &=& M^2 P(r),  \\
\trho(\tr) &=& M^2 \rho(r),  \\
 \tilde{\nu}(\tr) &=& \nu(r).
\end{eqnarray}
Besides, we also let $\tR \equiv R/M$ and $\tX \equiv R_*/M$,
where $R_*=r_{*}(r=R)$ is the tortoise radius of the star.

\begin{figure}
\includegraphics[angle=270,width=8.5cm]{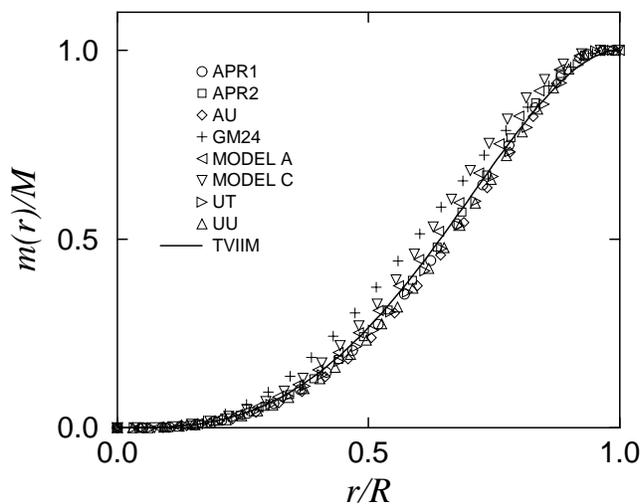}
\caption{The mass distributions of various realistic neutron stars
and TVIIM with a common compactness $\cc=0.2$ are plotted against
$r/R$.} \label{f2}
\end{figure}

From (\ref{KG_sc}) it is then apparent that the mass-independence
of the scaled frequency $\tomega$ follows directly from that of
the scaled potential $\tV$. Motivated by this and noting that the
following mass distribution:
\begin{equation}
m_\c(r)= M \left[ \frac{5}{2}\left(\frac{r}{R}\right)^3
-\frac{3}{2}\left(\frac{r}{R}\right)^5 \right],
\end{equation}
conventionally termed as TVIIM in the literature
\citep{Tolman:1939jz,Lattimer:2001}, is indeed a good
approximation of $m(r)$ for neutron stars with various EOSs (see
Fig.~\ref{f2}), we demonstrated that QNMs of TVIIM (solid circles
in Fig.~\ref{f1}) display universal behavior observed in realistic
stars and interpreted such universality
 as the consequence of the mass-independence of the
scaled mass distribution of TVIIM \citep{preprint}:
\begin{equation}
\tm_\c(\tr)\equiv \frac {m_\c(r)}{M}=\frac{5}{2}\cc^3\tr^3
-\frac{3}{2}\cc^5\tr^5.
\end{equation}

\begin{figure}
\includegraphics[angle=270,width=8.5cm]{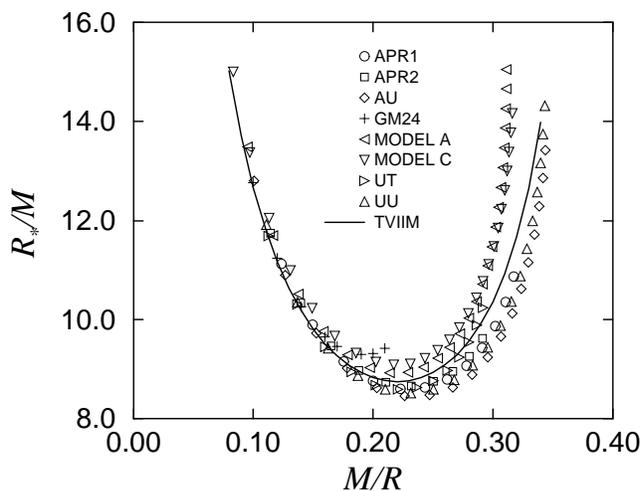}
\caption{The figure shows the relationship between $R_*/M$ and the
compactness for TVIIM (solid line) and other realistic neutron
stars.}\label{f3}
\end{figure}

In addition to approximating the mass distribution in realistic
neutron stars, TVIIM is a solvable model whose metric coefficients
$e^{\lambda}$ and $e^{\nu}$ can be obtained in closed forms
\citep{Tolman:1939jz}:
\begin{eqnarray}
e^{-\lambda}&=&1-{\cal C} \xi^2(5-3\xi^2)\ , \label{e_lam_eq}\\
e^{\nu}&=&(1-5{\cal C}/3)\cos^2\phi \ . \label{e_phi_eq}
\end{eqnarray}
Here $\xi=r/R$,
\begin{eqnarray}
\phi&=&({w}_1-{w})/2+\phi_1 \ , \\
{w}&=&\log\left[\xi^2-\frac{5}{6}+\sqrt{\frac{e^{-\lambda}}{3\cal{C}}}\right],\\
\phi_1&=&\phi(\xi^2=1)=\arctan\sqrt{\frac{\cal{C}}{3(1-2\cal{C})}}\ ,\\
w_1&=&w(\xi^2=1)\ ,
\end{eqnarray}
and the pressure is given by:
\begin{eqnarray}
 P=\frac{1}{4\pi
R^2}\left[\sqrt{3\cc
e^{-\lambda}}\tan\phi-\frac{\cc}{2}(5-3\xi^2)\right],
\end{eqnarray}
By using (\ref{e_lam_eq}), (\ref{e_phi_eq}) and the definition of
$r_*$, we obtain
\begin{eqnarray}\label{intI_eq}
    r_{*}&=&R\int^{\xi}_0
     I(\xi',\cc) \d \xi',
\end{eqnarray}
where
\begin{equation}
I(\xi,\cc)=\left\{\left[1-{\cal C} \xi^2(5-3\xi^2)\right](1-5{\cal
C}/3)\cos^2\phi \right\}^{-1/2}.
\end{equation}
Hence, the scaled tortoise radius of the star is
\begin{eqnarray}\label{STR}
\tX = \frac{1}{\cc}\int^{1}_0 I(\xi,\cc) \d \xi.
\end{eqnarray}
Since TVIIM has a  mass profile similar to those of realistic
neutron stars, we expect differences in $\tX$ between TVIIM and
realistic neutron stars would be small. In fact, our conjecture is
verified by Fig.~\ref{f3} where  $R_*/M$ is plotted against $M/R$
for TVIIM and other realistic stars.

As the integral in (\ref{intI_eq}) could not be evaluated
analytically, we are not able to express $V(r_*)$ in terms of
simple functions of $r_*$ and hence it is not possible to find the
analytical form of the wave function in NSRWE even inside the
star. However, we could expand $V(r_{*})$ as a power series around
the center of the star:
\begin{equation}\label{VCEA}
    \tilde{V}(\tx)=\frac{l(l+1)}{\tx^2}+\tilde{\alpha}+O(\tx^2),
\end{equation}
where $\tilde{\alpha}$ is a constant dependent on $\cc$ and $l$.
It is worthwhile to note that the leading term in the expansion of
$\tilde{V}(\tilde{r}_*)$ is nothing but a centrifugal barrier in
$\tilde{r}_*$.

Despite that the leading two terms of the expansion in
(\ref{VCEA}) can nicely approximate the potential at the center of
the star, it usually produces a discontinuity across the stellar
surface. As QNMs are sensitive to discontinuities, we have to
adopt another potential $\tilde{V}_{\c}(\tx)$ given by
\begin{equation}\label{CBAV}
    \tilde{V}_{\c}(\tx)=\frac{l(l+1)}{\tx^2}-\frac{l(l+1)}{\tX^2}+\left(1-\frac{2}{\tR}\right)
    \left(\frac{l(l+1)}{\tR^2}-\frac{6}{\tR^3}\right), \quad \tx
    \le \tX ,
\end{equation}
where $R_*$ is obtained from (\ref{STR}). This potential is
similar to the one in (\ref{VCEA}) around $\tilde{r}_*=0$ and, in
addition, is continuous across the stellar surface. This scheme is
termed the centrifugal barrier approximation (CBA) in our paper.
It is worthwhile to note that under CBA the wave function of NSRWE
inside the star is simply given by $\tilde{k}\tx
j_l(\tilde{k}\tx)$, e.g. for $l=2$:
\begin{equation}
\tpsi_{}(\tx)=\frac{3\sin(\tilde{k}\tx)}{\tilde{k}^2\tx^2}-\frac{3\cos(\tilde{k}
\tx)}{\tilde{k}\tx}-\sin(\tilde{k}\tx),\quad \tx
    \le \tX.
\end{equation}
Here
\begin{equation}\label{}
\tilde{k}=\left[\tomega^2+\frac{l(l+1)}{\tX^2}-\left(1-\frac{2}{\tR}\right)
    \left(\frac{l(l+1)}{\tR^2}-\frac{6}{\tR^3}\right)\right]^{1/2}
\end{equation}
is the scaled effective  wave number.

In addition to its simplicity, as shown in Fig.~\ref{f1}, CBA also
yields QNMs (denoted by the solid line in the figure) close to
those of TVIIM and realistic neutron stars. Therefore, QNMs of CBA
manifest the universal behavior (\ref{TLA}). This remarkable
discovery is indeed the crux of the present paper. As the scaled
potential in (\ref{CBAV}) is completely characterized by
$\tR=1/\cc$ and $\tX$, which is also a function of $\cc$, the
cause of the observed universality in neutron star axial
pulsations then becomes obvious and understandable. In a nutshell,
the universality is attributable to (i) the centrifugal potential
at the center of the star; (ii) the continuity of the potential;
and (iii) the tortoise radius $\tX$ of TVIIM star.
\section{Scaled-Coordinate Logarithmic Perturbation}
After locating the physical origin of the universality in QNMs of
neutron stars, we further consider the following questions: (i)
Can the coefficients $a$, $b$ and $c$ in (\ref{TLA}) be obtained
analytically from CBA? (ii) Can QNMs of realistic stars be
obtained from those of CBA? We hold positive views on these
challenging issues. In this paper, we aim to develop a
perturbative study to consider how changes in compactness and EOS
 could affect the frequencies of the QNMs of a star.
 In fact, the simplicity and generality of CBA
directly leads to a feasible perturbative analysis for QNMs of
neutron stars and in turn provides appropriate solutions to these
two questions.

As discussed above, axial oscillations of neutron stars are
described by NSRWE with a potential term dependent on the
distribution of energy-mass density and the pressure inside a
star. QNMs of these oscillations are the eigen-solution to
(\ref{KG_eq}) and, in addition, they are regular at the origin and
satisfy the outgoing boundary condition at spatial infinity.  It
is well known that the wave function of QNMs diverges at spatial
infinity and is not amenable to standard perturbation theory
\citep[see, e.g.][]{dirty2}.  To this end, we will generalize the
logarithmic perturbation theory for QNMs, previously formulated by
\citet{dirty1,dirty2} to study how the QNMs of a black hole
respond to static perturbations such as a static mass shell, to
consider QNMs of neutron stars. In this case, the scaled NSRWE
(\ref{KG_sc}) that makes use of the scaled coordinate $\tr$ is
obviously more amenable to perturbative expansion since outside
the star $\tV(\tr)$ depends only $\tr$ and is independent of the
stellar mass. Therefore, the method developed here is referred to
as the Scaled-Coordinate Logarithmic Perturbation Theory (SCLPT).
In the subsequent discussion, we will show that shifts in scaled
QNM frequencies can be expressed in terms of integrals with finite
domains of integration.

\subsection{Perturbative expansion}
To formulate a perturbative expansion for the QNM frequency, we
introduce a formal perturbation parameter, $\mu$, to measure the
departure of the stellar configuration from an unperturbed one,
which has a scaled circumferential (tortoise) radius
$\tilde{R}_{0}$ ($\tilde{R}_{*0}$), a scaled potential
$\tilde{V}_0(\tilde{r}_{*})$, and a QNM with a  scaled frequency
$\tomega_0$. Analogous to other perturbation theories, we will
assume that the QNM wave function is known. This assumption does
not pose any problems to our current study because we will use the
CBA potential as the unperturbed system and, as mentioned above,
the wave function can be obtained analytically. Besides,
$\tilde{R}_{*0}$  and $\tilde{V}_0(\tilde{r}_{*})$ can be obtained
from (\ref{STR}) and (\ref{CBAV}), respectively. The subject of
interest in this paper is to obtain the QNM frequency of a
perturbed star  whose potential function, as a function of $\tx$,
can be expressed as:
\begin{eqnarray}\label{Vexp}
\tV(\tx)&=&\tilde{V}_0(\tx)+\Delta \tV \nonumber \\
&=&\tilde{V}_0(\tx)+\mu\tilde{V}_1(\tx)+\mu^2\tilde{V}_2(\tx)+...~,
\,
\end{eqnarray}
for $\tx \leq \tX(\mu)$. Here $\tilde{R}(\mu)$ and
$\tilde{R}_{*}(\mu)$ are the circumferential and tortoise radii of
the perturbed star, respectively, which are also functions of
$\mu$, with $\tilde{R}(\mu=0)=\tilde{R}_{0}$ and
$\tilde{R}_*(\mu=0)=\tilde{R}_{*0}$.

Outside the star, it is clear from (\ref{SP}) that $\tV$, as a
function of $\tilde{r}$, is independent of $\mu$ and hence
$\Delta\tV=0$ for $\tr \geq \tR(\mu)$.  It is therefore more
convenient to adopt the scaled tortoise coordinate $\tx$  and the
scaled circumferential radius $\tr$ to describe wave propagation
inside and outside the star, respectively. Each of these radial
coordinates has its own advantage. The scaled tortoise coordinate
description casts the NSRWE into the standard Klein-Gordon
equation form, whereas the use of the scaled circumferential
radius simplifies the form of $\tV$ outside the star.  In our
formalism, we will apply these coordinates  in different regions
of interest.

In the following we seek a power series expansion for the QNM
frequency $\tomega(\mu)$ of the perturbed star:
\begin{equation}\label{Fexp}
\tomega(\mu)=\tomega_0+\mu\tomega_1+\mu^2\tomega_2+...~.
\end{equation}
We will find a general expression for $\tomega_n$ ($n=1,2,3,...$)
and explicitly determine the leading two expansion coefficients,
$\tomega_1$ and $\tomega_2$.

\subsection{Connection formula}
Inside the star we adopt the scaled tortoise coordinate $\tx$ as
the independent variable and consider the solution to
(\ref{KG_sc})  as a function of three independent variables $\tx$,
$\tomega$ and $\mu$, namely:
\begin{equation}
\tpsi=\tpsi_-(\tx,\tomega,\mu).
\end{equation}
This solution is specified by the regularity condition at the
origin and holds for $\tx < \tX$. Outside the star ($\tr \geq
\tR$), the scaled circumferential radius is used instead and we
regard the solution to (\ref{KG_sc}) there as a function of $\tr$
and $\tomega$. In particular, $\tpsi_+(\tr,\tomega)$ is a solution
that satisfies the outgoing wave boundary condition at spatial
infinity. Outside the star, NSRWE is identical to the
Regge-Wheeler equation of black holes and therefore
$\tpsi_+(\tr,\tomega)$  can be obtained from the Leaver's series
solution originally developed for determination of QNMs of black
holes \citep{Leaver_series,Liu,dirty2}.

For each $\mu$, we can evaluate the QNM frequency $\tomega$ by
matching the logarithmic derivatives of $\tpsi_+$ and $\tpsi_-$ at
$\tr=\tR(\mu)=\tR_0+\mu\tR_1+\mu^{2}\tR_{2}+\cdots$, i.e.
\begin{equation}\label{connect}
f_-(\tX(\mu),\tomega(\mu),\mu) = f_+(\tR(\mu),\tomega(\mu)),
\end{equation}
where
\begin{eqnarray}
f_-(\tx,\tomega,\mu)&=&\frac{1}{\tpsi_-(\tx,\tomega,\mu)}\frac{\partial\tpsi_-}{\partial\tx}, \\
f_+(\tr,\tomega)&=&\frac{1}{\tpsi_+(\tr,\tomega)}\frac{\partial\tpsi_+}{\partial\tx},
\nonumber \\
&=&\frac{(\tr-2)}{\tr\tpsi_+(\tr,\tomega)}\frac{\partial\tpsi_+}{\partial\tr}
.
\end{eqnarray}
It is interesting to note that $f_{\pm}$ are matched at the scaled
radius $\tR(\mu)$ (or $\tX(\mu)$) that would vary with $\mu$.

\subsection{Expansion of logarithmic derivative}
To generate an expansion for the QNM frequency $\tomega(\mu)$, we
firstly expand $f_-$ as a power series in the perturbation
parameter $\mu$:
\begin{eqnarray}
f_-(\tx,\tomega(\mu),\mu)&=&f_0(\tx)+\mu f_1(\tx)+\mu^2
f_2(\tx)+...~,\label{fminus}
\end{eqnarray}
where $f_-(\tx,\tomega_0,\mu=0)=f_0(\tx)$ and we have suppressed
the dependence on $\tomega_0$ in the expansion coefficients
$f_i(\tx)$ ($i=1,2,3,\ldots$). By transforming (\ref{KG_sc}) into
the form of the Riccati equation:
\begin{equation}\label{riccati_eqn}
f_-'+f_-^2+\tomega^2- \tilde{V}(\tx)=0,
\end{equation}
making use of the expansions (\ref{Vexp}) and (\ref{Fexp}), and
comparing equal powers of $\mu$, we show that
\begin{equation}\label{riccati_eqn_2}
f_n'+2f_0f_n+2\tomega_0\tomega_n=\tU_n, \quad n=1,2,...\,.
\end{equation}
Here $\tU_1(\tx)=\tV_1(\tx)$ and for $n \ge 2$,
\begin{equation}
\tU_n(\tx)=\tV_n(\tx)-\sum_{i=1}^{n-1}[f_i(\tx)f_{n-i}(\tx)+\tomega_i\tomega_{n-i}],
\end{equation}
and we have used the prime to symbolize differentiation with
respect to $\tx$. Eq.~(\ref{riccati_eqn_2}) can be solved by
introducing the integrating factor
$\exp[2\int^{\tx}dyf_0(y)]=\tpsi_0^2(\tx)$, resulting in an
integral expression for $f_n(\tilde{R}_{*0})$:
\begin{equation}
f_n(\tilde{R}_{*0})\tpsi_0^2(\tilde{R}_{*0})=\int_{0}^{\tilde{R}_{*0}}\d\tx[\tU_n(\tx)-2\tomega_0\tomega_n]\tpsi_0^2(\tx).
\end{equation}

Secondly, as the logarithmic derivatives are to be matched at
$\tx=\tX(\mu)=\tR_{*0}+\mu\tR_{*1}+\mu^{2}\tR_{*2}+\cdots$, we
expand $f_-(\tX(\mu),\tomega(\mu),\mu)$ and
$f_+(\tR(\mu),\tomega(\mu))$ in power series of $\mu$, yielding:
\begin{eqnarray}\label{fminusmu}
f_-(\tX(\mu),\tomega(\mu),\mu)&=&f_0(\tX(\mu))+\mu
f_1(\tX(\mu))+\mu^2f_2(\tX(\mu))+...\nonumber\\
&=&\left[f_0(\tilde{R}_{*0})+\mu\frac{\d\f_0}{\d\mu}+\frac{\mu^2}{2!}\frac{\d^2f_0}{\d\mu^2}+...\right]\nonumber\\
&&+\mu\left[f_1(\tilde{R}_{*0})+\mu\frac{\d f_1}{\d\mu}+\frac{\mu^2}{2!}\frac{\d^2f_1}{\d\mu^2}+...\right]\nonumber\\
&&+\mu^2\left[f_2(\tilde{R}_{*0})+\mu\frac{\d
f_2}{\d\mu}+\frac{\mu^2}{2!}\frac{\d^2f_2}{\d\mu^2}+...\right]
+...~,
\end{eqnarray}
and
\begin{eqnarray}\label{fplusmu}
f_+(\tR(\mu),\tomega(\mu))&=&f_+(\tR(\mu),\tomega(\mu))\nonumber\\
&=&f_+(\tR_0,\tomega_0)+\mu\frac{\d
f_+}{\d\mu}+\frac{\mu^2}{2!}\frac{\d^2f_+}{\d\mu^2}+...~.
\end{eqnarray}
Here it is understood that all derivatives with respective to
$\mu$ are evaluated at $\mu=0$. Hence, by comparing equal powers
of $\mu$ in (\ref{fminusmu}) and (\ref{fplusmu}),
$f_n(\tilde{R}_{*0})$ can be expressed in terms of derivatives of
$f_{i}$ ($i<n$) and $f_+$ as follows:
\begin{eqnarray}
f_n(\tilde{R}_{*0})&=&\frac{1}{n!}\frac{\d^nf_+}{\d\mu^n}-\left[\frac{1}{n!}\frac{\d^nf_0}{\d\mu^n}
+\frac{1}{(n-1)!}\frac{\d^{n-1}f_1}{\d\mu^{n-1}}+\cdots+\frac{\d f_{n-1}}{\d\mu}\right]\nonumber\\
&=&\frac{1}{n!}\frac{\d^nf_+}{\d\mu^n}-\Theta_n,
\end{eqnarray}
where, by definition, $\Theta_n$ is equal to the sum in the square
bracket. Furthermore, by introducing another expression $\Delta_n$
defined by:
\begin{equation}\label{}
\Delta_n=
\left[f_+(\tR_0+\mu\tR_1+\cdots+\mu^{n-1}\tR_{n-1},\tomega_0+\mu\tomega_1
+\cdots+\mu^{n-1}\tomega_{n-1})-f_+(\tR_0,\tomega_0)\right]_n
\end{equation}
where $[F(\mu)]_n$ indicates the $n$th-order term of a function
$F(\mu)$, it is readily shown that
\begin{eqnarray}
\frac{1}{n!}\frac{\d^nf_+}{\d\mu^n}&=
&\Delta_n+\left(\tomega_n\frac{\partial}{\partial\tomega_0}
+\tR_n\frac{\partial}{\partial\tr}\right)f_+(\tomega_0,\tR_0).
\end{eqnarray}
As a result, we obtain a formal expansion for the scaled frequency
$\tomega$, which reads:
\begin{equation}\label{freq1}
\tomega_n=\frac{\langle\tpsi_0|\tU_n|\tpsi_0\rangle}{2\tomega_0\langle\tpsi_0|\tpsi_0\rangle},
\end{equation}
with
\begin{equation}\label{freq2}
\langle\tpsi_0|\tU_n|\tpsi_0\rangle=\int_{0}^{\tilde{R}_{*0}}\d\tx\tU_n(\tx)\tpsi_0^2(\tx)
+\left(\Theta_n-\Delta_n-\tR_n\frac{\partial
f_+}{\partial\tr}\right)\tpsi_0^2(\tilde{R}_{*0}),
\end{equation}
and
\begin{equation}\label{freq3}
\langle\tpsi_0|\tpsi_0\rangle=\int_0^{\tilde{R}_{*0}}\tpsi_0^2(\tx)\d\tx+\frac{\tpsi_0^2(\tilde{R}_{*0})}{2\tomega_0}
\frac{\partial f_+}{\partial\tomega_0}.
\end{equation}

Despite the simplicity of (\ref{freq1}), which is akin to standard
perturbation formulas in quantum mechanics with
$\langle\tpsi_0|\tpsi_0\rangle$ playing the role of the norm
squared of a quantum state, the emergence of terms like $\tU_n$,
$\Delta_n$ and $\Theta_n$ reveals the achievement underlying this
formula. It is worthy of remark that in (\ref{freq2}) there are
three different contributions to the frequency shift $\tomega_n$,
namely an integral over the interior of the star and two surface
terms originating from $f_+$ and $f_-$, respectively. In the
following discussion, we will work out explicit expressions for
the first and second order results.
\subsection{First and second order frequency shifts}
For the case $n=1$, the expression of $\tU_1(\tx)$ is trivial,
while $\Theta_n$ and $\Delta_n$ are given by:
\begin{eqnarray}
\Theta_1&=&\frac{\d f_0}{\d\mu}=\frac{\d f_0}{\d\tx}\frac{\d\tX}{\d\mu} \\
\Delta_1&=&0.
\end{eqnarray}
As a result, it is clear that
\begin{equation}\label{U_1}
\langle\tpsi_0|\tU_1|\tpsi_0\rangle=\int_{0}^{\tilde{R}_{*0}}\d\tx\tV_1(\tx)\tpsi_0^2(\tx)
+\left(\tR_{*1}\frac{d f_0}{d\tx}-\tR_1\frac{\partial
f_+}{\partial\tr}\right)\tpsi_0^2(\tilde{R}_{*0}),
\end{equation}
and hence $\tomega_1$ can be readily obtained from (\ref{freq1})
and (\ref{freq3}).

For the case $n=2$, it is straightforward to show that
\begin{eqnarray}
\tU_2(\tx)&=&\tV_2(\tx)-f_1^2(\tx)-\tomega_1^2\;;\\
\Theta_2&=&\frac{1}{2}\frac{\d^2f_0}{\d\mu^2}+\frac{\d f_1}{\d\mu}
\nonumber\\
&=&\frac{\d f_0}{\d\tx}\tR_{*2}+\frac{1}{2}\frac{\d^2f_0}{\d\tx^2}
\tR_{*1}^2+\frac{\d f_1}{\d\tx}\tR_{*1}\;;\\
\Delta_2&=&\left[f_+(\tR_0+\mu\tR_1,\tomega_0+\mu\tomega_1)-
f_+(\tR_0,\tomega_0)\right]_2\nonumber\\
&=&\frac{1}{2}\tomega_1^2\frac{\partial^2f_+}{\partial\tomega^2}+
\tR_1\tomega_1\frac{\partial
f_+}{\partial\tr\partial\tomega}
+\frac{1}{2}\tR_1^2\frac{\partial^2f_+}{\partial\tr^2}.
\end{eqnarray}
Direct substitution of these results into (\ref{freq2}) leads to
the second-order term:
\begin{eqnarray}\label{U_2}
&&\langle\tpsi_0|\tU_2|\tpsi_0\rangle \nonumber
\\&=&\int^{\tR_{*0}}_0\left[\tV_2(\tx)-f^2_1(\tx)-\tomega_1^2\right]\tpsi^2_0(\tx)\d\tx
\nonumber \\&&+\tpsi^2_0(\tR_{*0})\left[\frac{\d
f_0}{\d\tx}\tR_{*2}+\frac{\tR_{*1}^2}{2}
\frac{\d^2 f_0}{\d\tx^2}+\frac{\d f_1}{\d\tx}\tR_{*1}\right]_{\tx=\tR_{*0}}\nonumber\\
&&-\tpsi^2_0(\tR_{*0})\left[\frac{\tomega_1^2}{2}\frac{\partial^2f_+}
{\partial\tomega^2}+\tR_1\tomega_1\frac{\partial
f_+}{\partial\tr\partial\tomega}+\frac{\tR_1^2}{2}\frac{\partial^2f_+}{\partial\tr^2}
+\tR_2\frac{\partial f_+}{\partial\tr}\right]_{\tr=\tR_0}\,,
\end{eqnarray}
and the second-order frequency change follows directly from
(\ref{freq1}).

After succeeding in deriving the first and second order shifts in
the eigenfrequency of QNMs of neutron stars, we will apply
relevant formulas to answer the questions posed in the beginning
of this section.

\section{Universality in QNMs}
Being a good global approximation to realistic stars, CBA also
demonstrates the  universal behavior summarized by (\ref{TLA}). In
fact, the solid line in Fig.~\ref{f1}, representing  the QNM
frequencies of CBA, is close to the best quadratic fit to those of
realistic stars (the dotted line). Therefore, we expect that the
universal behavior displayed by realistic neutron stars can be
understood from the QNMs of CBA and Eq.~(\ref{TLA}) can be deduced
from CBA as well. Motivated by this conjecture, we evaluate the
QNM frequency of CBA with SCLPT.

In SCLPT, the QNM frequencies for CBA stars with different
compactness can be obtained by considering compactness $\cc$ as
the perturbation parameter. In this case, the formal expansion
parameter $\mu=\cc-\cc_0$, where $\cc_0$ is the compactness of a
reference CBA star whose QNMs are known. Hence, to second order in
$\cc-\cc_0$, $\tomega(\cc)$ is approximately given by:
\begin{equation}\label{f_exp}
\tomega(\mu)=\tomega_0+(\cc-\cc_0)\tomega_1+(\cc-\cc_0)^2\tomega_2
\,,
\end{equation}
where $\tomega_0$ is the QNM frequency of the reference CBA star.
The first and second order shifts in QNM frequencies are
proportional to $\cc-\cc_0$ and $(\cc-\cc_0)^2$, respectively.
Under this approximation, the QNM frequency is expressible in
terms of a quadratic function of $\cc$:
    \begin{equation}\label{omega_quad}
    \tomega=a\cc^2+b\cc+c\,,
    \end{equation}
with
\begin{eqnarray}
a&=&\tomega_2;\label{a}\\
b&=&\tomega_1-2\cc_0\tomega_2;\label{b} \\
c&=&\tomega_0-\cc_0\tomega_1+\cc_0^2\tomega_2.\label{c}
\end{eqnarray}
It is obvious that Eq.~(\ref{omega_quad}) is in prefect agreement
with the universal behavior (\ref{TLA}) discovered numerically by
\citet{Andersson1998,Ferrari,preprint}.

\begin{figure}
\includegraphics[angle=270,width=8.5cm]{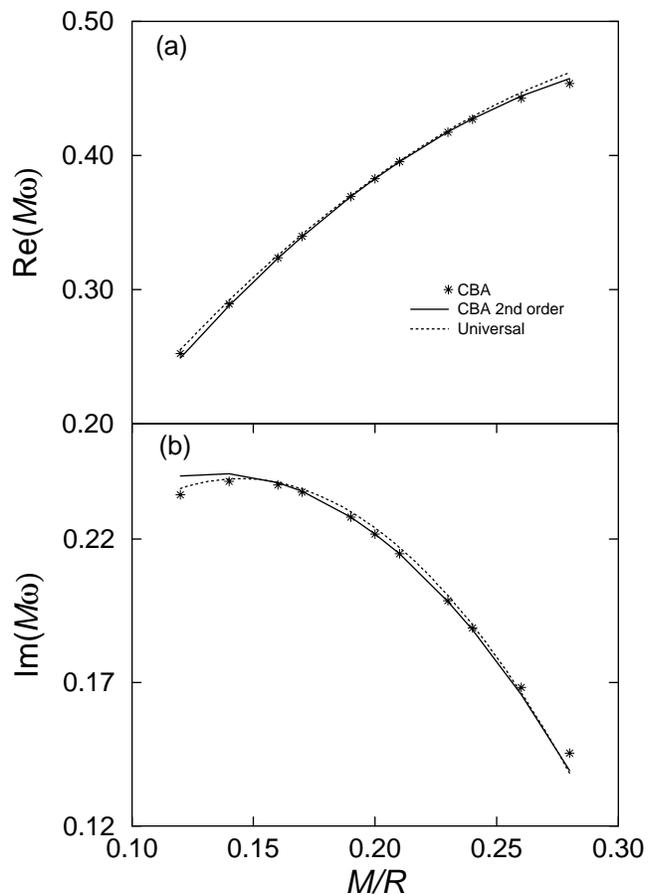}
\caption{The scaled QNM frequency of the CBA star is plotted
against the compactness $\cc$. The stars and the solid line
represent the exact values and the ones obtained from applying the
second-order SCLPT to a CBA star with $\cc=0.2$, respectively. For
purpose of comparison, we also show the universal curve (the
dotted line), i.e. the best quadratic fit to QNM frequencies of
the realistic stars considered in this paper.} \label{f6}
\end{figure}

To gauge the accuracy of the second-order SCLPT mentioned above,
we apply it to a reference CBA star with compactness $M/R=0.2$. As
shown in Fig.~\ref{f6}, the results obtained from SCLPT
(represented by the solid line) are good approximation of the
exact QNMs (represented by the stars). This clearly demonstrates
the validity of SCLPT. In addition, the perturbative results also
faithfully demonstrate the universal behavior  displayed by
realistic stars (represented by the dotted line), and the
numerical values of $a$, $b$ and $c$ obtained from
(\ref{a}),(\ref{a}) and (\ref{a}) respectively are in nice
agreement with the those obtained from the best quadratic fit to
   the QNMs of the realistic stars (see Table~1 for reference).
Hence, the universality in axial pulsations of neutron stars is
fully understood and predicted analytically.

    The technical details of the perturbation scheme  yielding
    $\tomega_1$, $\tomega_2$ and hence
    the constants $a$, $b$ and $c$ are as follows. To obtain the
    first and second order shifts, we have to evaluate the
    all quantities appearing in (\ref{U_1}) and (\ref{U_2}).
    Specifically, $\tilde{R}_1$ and $\tilde{R}_2$ are given by:
\begin{eqnarray}\label{}
\tR_1&=&\left(\frac{\d\tR}{\d\cc}\right)_{\cc=\cc_0}=-\frac{1}{\cc_0^2},
\\
\tR_2&=&\frac{1}{2}\left(\frac{\d^2\tR}{\d\cc^2}\right)_{\cc=\cc_0}=\frac{1}{\cc_0^3}.
\end{eqnarray}
Analogously, $\tR_{*1}$ and $\tR_{*2}$ can be obtained:
\begin{eqnarray}\label{}
\tR_{*1}&=&\left(\frac{\d\tX}{\d\cc}\right)_{\cc=\cc_0}\nonumber
\\
&=&\frac{1}{\cc_0}\int^{1}_0 \left[\frac{\partial
I(\xi,\cc_0)}{\partial
\cc_0 }-\frac{I(\xi,\cc_0)}{\cc_0}\right] \d\xi , \\
\tR_{*2}&=&
\frac{1}{2}\left(\frac{\d^2\tX}{\d\cc^2}\right)_{\cc=\cc_0}
\nonumber
\\ &=& \frac{1}{2\cc_0}\int^{1}_0 \left[\frac{\partial^2
I(\xi,\cc_0)}{\partial \cc_0^2 }-\frac{2}{\cc_0}\frac{\partial
I(\xi,\cc_0)}{\partial \cc_0
}+\frac{2I(\xi,\cc_0)}{\cc_0^2}\right] \d\xi .
\end{eqnarray}
On the other hand, as $\tilde{V}_c$ contains explicit dependence
on $\tX$ and $\tR$, $\tV_1$ and $\tV_2$ can be found:
\begin{eqnarray}\label{}
\tV_1(\tx)&=&\left(\frac{\partial\tV_c}{\partial\cc}\right)_{\cc=\cc_0}\nonumber
\\
&=& \frac{2l(l+1)\tR_{*1}}{\tR_{*0}^3}+
2l(l+1)\cc_0-3(l^2+l+3)\cc_0^2+12\cc_0^3 \nonumber
\\ \tV_2(\tx)&=&\left(\frac{\partial^2\tV_c}{\partial\cc^2}\right)_{\cc=\cc_0}\nonumber
,\\
&=& -\frac{6l(l+1)\tR_{*1}^2}{\tR_{*0}^4}+\frac{4
l(l+1)\tR_{*2}}{\tR_{*0}^3}+ 2l(l+1)-6(l^2+l+3)\cc_0+36\cc_0^2,
\end{eqnarray}
which are constants independent of $\tx$.

It is easy to see that $\tV_n(\tx)$ ($n=1,2,3,\ldots$) are indeed
all $\tx$-independent under CBA. This fact greatly simplifies the
perturbation calculation in CBA and renders the integrals involved
in the evaluation of $\tomega_1$ and $\tomega_2$ exactly solvable.
For example, noting that $\tpsi_0(\tx)=\tilde{k}\tx
j_l(\tilde{k}\tx)$, we have
\begin{equation}\label{}
\int_{0}^{\tilde{R}_{*0}}\d\tx\tV_1(\tx)\tpsi_0^2(\tx) =
\frac{\tV_1
\tR_{*0}}{2}\left\{x^2\left[j_l^2(x)+j_{l-1}^2(x)\right]-
(2l+1)xj_l(x)j_{l-1}(x)\right\}_{x=\tilde{k} \tR_{*0}}
\end{equation}
and
\begin{eqnarray}
f_1(\tilde{r}_{*})&=&\frac{1}{\tpsi_0^2(\tilde{r}_{*})}
\int_{0}^{\tilde{r}_{*}}\d\tx'[\tV_1(\tx')-2\tomega_0\tomega_1]
\tpsi_0^2(\tx') \nonumber \\
&=&\frac{(\tV_1-2\tomega_0\tomega_1)\tR_{*0}}{2}\left[1+\frac{j_{l-1}^2(x)}{j_l^2(x)}-
\frac{(2l+1)j_{l-1}(x)}{xj_l(x)}\right]_{x=\tilde{k} \tR_{*0}}\,.
\end{eqnarray}

\begin{figure}
\includegraphics[angle=270,width=8.5cm]{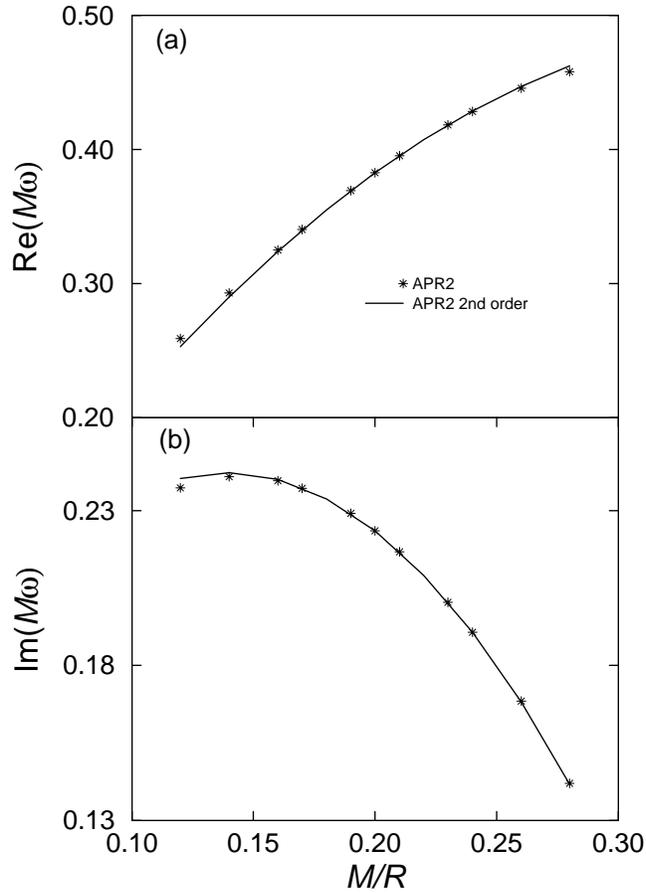}
\caption{The scaled QNM frequency of the APR2 star is plotted
against the compactness $\cc$. The stars and the solid line are
the exact values and the ones obtained by applying the
second-order SCLPT to an APR2 star with $\cc=0.2$, respectively.}
\label{f7}
\end{figure}

Besides, we have also apply SCLPT to evaluate QNMs of other
realistic stars. As an example, we show here the result for a star
constructed with APR2 EOS. As demonstrated in Fig.~\ref{f7},  the
second order result of SCLPT is again a good approximation to the
exact numerical results. This provides independent corroboration
to the validity of SCLPT.

\section{Individuality of QNMs}
\begin{figure}
\includegraphics[angle=270,width=8.5cm]{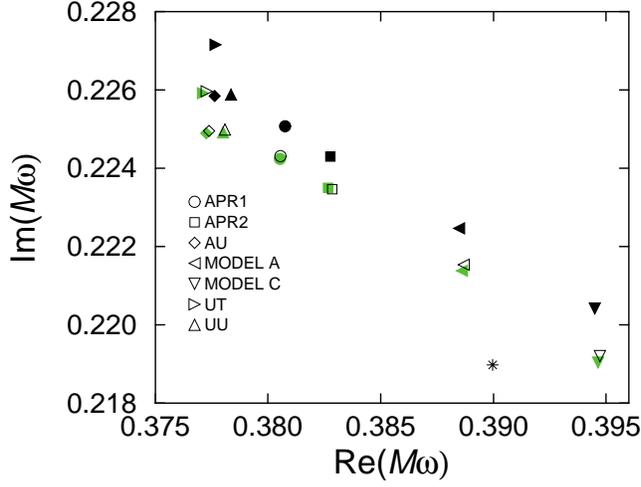}
 \caption{The star symbol shows the scaled QNM frequency of a CBA star with compactness
 $\cc= 0.2$.
 By applying SCLPT to the CBA star, the first and second order
 results, respectively denoted by dark and grey symbols,
 for QNM frequencies of realistic
 stars with the same compactness are obtained. For comparison,
 exact numerical QNM frequencies of realistic
 stars are indicated by the corresponding unfilled symbols.}
\label{f4}
\end{figure}

After establishing the universality in axial pulsations of neutron
stars, we turn our attention to the individuality of such QNMs. As
clearly shown in Figs.~\ref{f1} and \ref{f4}, QNM frequencies of
individual realistic neutron stars in general deviate slightly
from the universal curve (\ref{TLA}). In the present paper we aim
to evaluate QNMs of realistic stars from those of CBA star with
SCLPT developed here. For a realistic star whose compactness $\cc$
and potential $\tV(\tx)$ are known, we compare it with a CBA star
with the same compactness $\cc$. Therefore, we have
$\tV_0=\tV_\c$, $\tV=\tV_\c+\tV_1$, and $\tX=\tR_{*0}+\tR_{*1}$,
where $\tR_{*0}$ is the tortoise radius of the CBA star. However,
in this case the scaled circumferential radii of the two stars in
consideration are the same. In other words, $\tR_n=0$ for
$n=1,2,\ldots$. Besides, it is convenient to take the formal
expansion parameter $\mu=1$ in the current situation. As both
$\tV_n=0$ and $\tR_{*n}=0$ for $n=2,3\ldots$, the matrix elements
$\langle\tpsi_0|\tU_1|\tpsi_0\rangle$ and
$\langle\tpsi_0|\tU_2|\tpsi_0\rangle$ can be simplified and
respectively take the form:
\begin{equation}
\langle\tpsi_0|\tU_1|\tpsi_0\rangle=
\int_{0}^{\tilde{R}_{*0}}\d\tx\tV_1(\tx)\tpsi_0^2(\tx)
+\tR_{*1}\frac{\d f_0}{\d\tx}\tpsi_0^2(\tilde{R}_{*0}),
\end{equation}
\begin{eqnarray}
&&\langle\tpsi_0|\tU_2|\tpsi_0\rangle\nonumber\\
&=&-\int^{\tR_{*0}}_0\left[f^2_1(\tx)+\tomega_1^2\right]\tpsi^2_0(\tx)\d\tx
\nonumber
\\&&+\tpsi^2_0(\tR_{*0})\left[\left(\frac{\tR_{*1}^2}{2} \frac{\d^2 f_0}{\d\tx^2}+
\tR_{*1}\frac{\d f_1}{\d\tx}\right)_{\tx=\tR_{*0}}
-\frac{\tomega_1^2}{2}\left(\frac{\partial^2f_+}
{\partial\tomega^2}\right)_{\tr=\tR_0}\right]\,.
\end{eqnarray}
The first and second order shifts in QNM frequency then follow
directly from (\ref{freq1}), (\ref{freq2}) and (\ref{freq3}).

Figure \ref{f4} shows QNMs obtained from the perturbation scheme
outlined above for stars constructed from various EOSs with a
common compactness $\cc=0.20$. The unfilled, dark, and grey
symbols respectively
 indicate the exact,  the first and the second order results,
 while the star is the unperturbed frequency of CBA. It is clearly
 shown that the second order perturbation results can nicely
 approximate the exact ones.  We have also applied SCLPT
 to  stable neutron stars with a larger (or smaller) compactness and found that
 the second-order results are indeed reliable. In fact, the first
 order formula readily suffices to yield accurate prediction of the
 shift in ${\rm Re}\,\tomega$.

\begin{figure}
\includegraphics[angle=270,width=8.5cm]{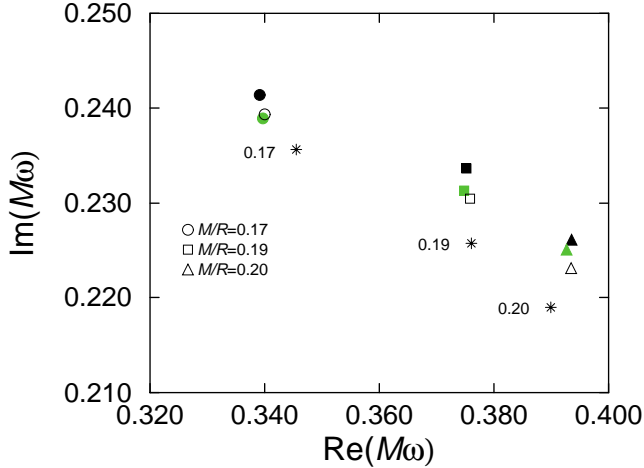}
\caption{The figure shows the scaled QNM frequencies of GM24 stars
with $\cc=0.17,0.19,0.20$. The star symbol indicates the QNM
frequencies of the corresponding CBA stars.
 The first and second order
 results obtained from SCLPT are denoted  by dark and grey symbols,
 respectively.
 For comparison
 exact numerical QNM frequencies of GM24
 stars are indicated by the corresponding unfilled symbols.} \label{f5}
\end{figure}

 On the other hand, we have to point out the limitations
of SCLPT. As in other cases, perturbation theory is likely to
break down in the presence of instabilities. Therefore, we expect
that the accuracy of perturbative results obtained from SCLPT gets
worse when the star becomes unstable against perturbation. We
verify this point by applying SCLPT to neutron stars constructed
with GM24 EOS \citep[][p.~244]{ComStar}, which are known to be
unstable if the compactness is greater than $0.21$. As clearly
demonstrated in Fig.~\ref{f5}, where QNMs of CBA and GM24 stars
with $\cc=0.17,0.19,0.20$ are shown, the deviation of the
second-order result from the exact value increases as the
compactness approaches the border of stability. However, as far as
${\rm Re}\,\tomega$ is concerned, both the first and second order
results are satisfactory.
\section{Conclusion and discussion}
The main achievement in the present paper is the discovery that
the universality in the QNM frequency of axial pulsations of
neutron stars in fact originates from the CBA, under which the
potential term in the scaled NSRWE is essentially a centrifugal
barrier. There are two parameters in the scaled potential, namely
the scaled circumferential radius $\tR$ and tortoise radius $\tX$
of the star. While the former is just the inverse of the
compactness $\cc$, the latter  can be determined from that of
TVIIM. These two parameters completely determine the QNMs of CBA
and in turn give rise to the observed universality.

In order to consider how the the physical characteristics of a
neutron star could affect the frequencies of its QNMs, we have
also developed a systematic perturbative scheme SCLPT, which is
designed to cope with the divergence in QNM wavefunction at
spatial infinity.
 Direct application of SCLPT to
CBA then successfully predicts the universality in the QNM
frequency of realistic neutron stars. The advantage of application
of the scaled coordinates ($\tr$ and $\tx$) in our study becomes
manifest in this regard because the potential $\tV$ outside a star
is independent of its detailed internal structure save its
compactness, leading to the observed universality
\citep{Andersson1998,Ferrari,preprint}.

On the other hand, SCLPT can also predict small deviations in QNM
frequencies from those of CBA for individual realistic neutron
stars. This opens possibilities for researchers to link the
 QNM frequencies observed from a distant star with its
internal structure and EOS as well. We are currently working on
the inverse problem of SCLPT, namely inferring the internal
structure of a pulsating neutron star from its gravitational wave
spectrum. Relevant progress in this direction will be reported
elsewhere in due course.

\begin{figure}
\includegraphics[angle=270,width=8.5cm]{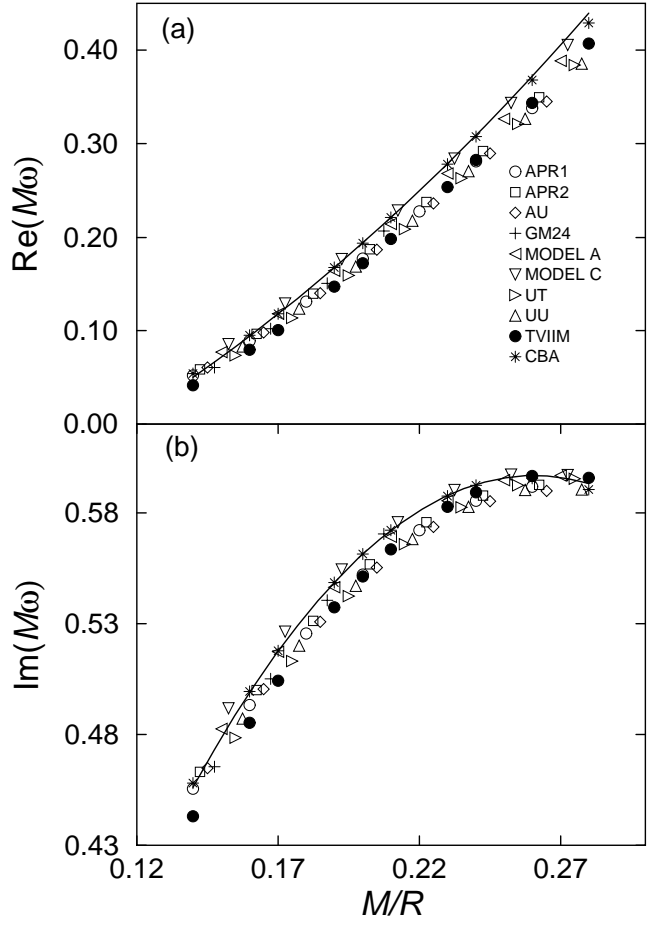}
\caption{The real and imaginary parts of the scaled QNM
frequencies of a $w_{\rm II}$-mode of various realistic stellar
models, TVIIM and CBA are plotted against the compactness. The
solid line shows the corresponding values obtained from the
second-order SCLPT.} \label{f8}
\end{figure}

\begin{figure}
\includegraphics[angle=270,width=8.5cm]{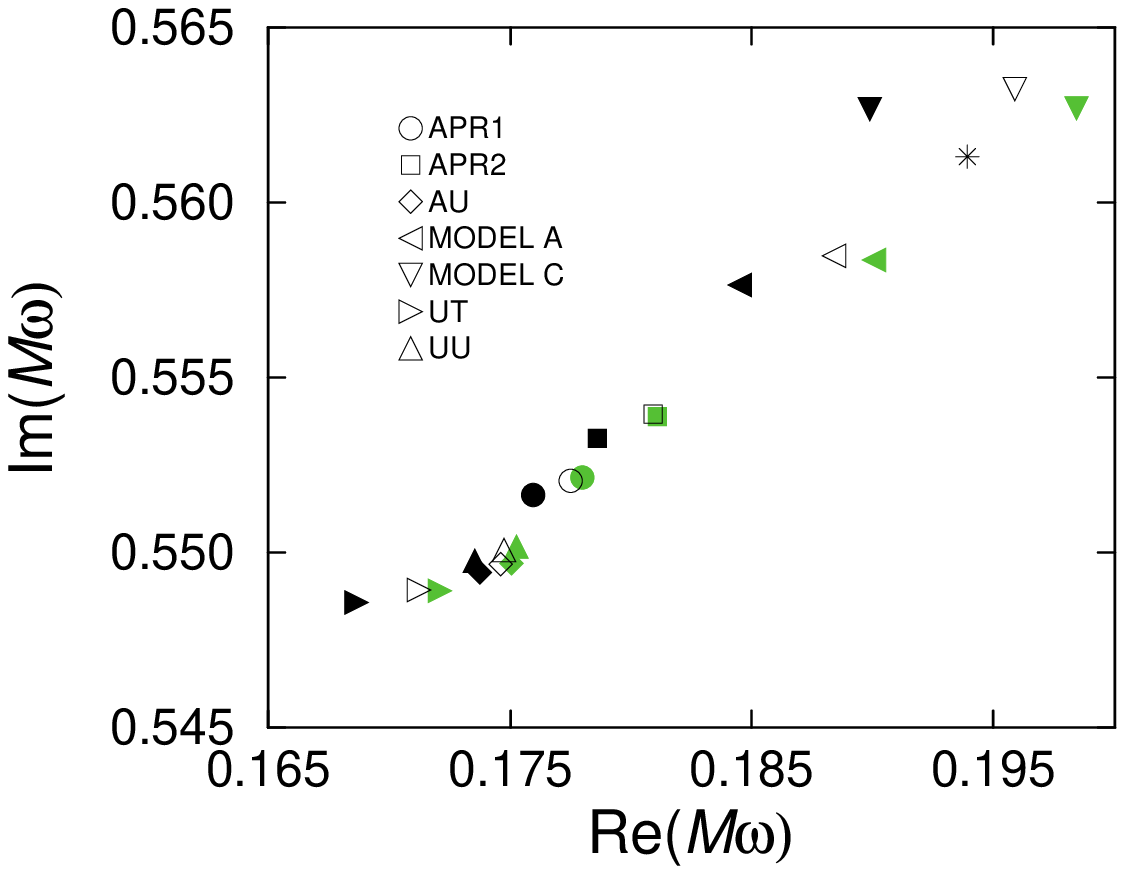}
 \caption{The star symbol shows the scaled QNM frequency of a $w_{\rm II}$-mode
 of a CBA star with compactness
 $\cc= 0.2$.
 By applying SCLPT to the CBA star, the first and second order
 results, respectively denoted by dark and grey symbols,
 for QNM frequencies of realistic
 stars with the same compactness are obtained. For comparison,
 exact numerical QNM frequencies of realistic
 stars are indicated by the corresponding unfilled symbols.}
\label{f9}
\end{figure}

So far we have used the least-damped $w$-mode to illustrate the
principle and accuracy of our method. However, the results
mentioned in the present paper are general and hold for modes of
higher orders (i.e. modes with frequencies higher than that of the
least-damped one). More interestingly, we note that the $w_{\rm
II}$-mode \citep[see e.g.][and references therein]{Kokkotas_rev}
also displays similar universality. As shown in Fig.~\ref{f8},
where the real and imaginary parts of the scaled QNM frequencies
of the $w_{\rm II}$-modes of various realistic stellar models,
TVIIM and CBA are plotted against the compactness, the
universality summarized in (\ref{TLA}) is clearly exhibited. It is
remarkable that the corresponding values obtained from application
of second-order SCLPT to CBA
 (the solid line) successfully capture the essence of
such universal behavior. To further examine the validity of CBA
and SCLPT in this case, we show in Fig.~\ref{f9} the scaled QNM
frequency of a $w_{\rm II}$-mode of a CBA star (the star symbol)
and other realistic stars (the unfilled symbols) with a common
compactness
 $\cc= 0.2$, the first and second order
 results obtained from applying SCLPT to the CBA star
 (respectively denoted by the corresponding dark and grey symbols). It
 is obvious that SCLPT can indeed reproduce accurate eigenfrequencies for
the $w_{\rm II}$-mode.

\begin{figure}
\includegraphics[angle=270,width=8.5cm]{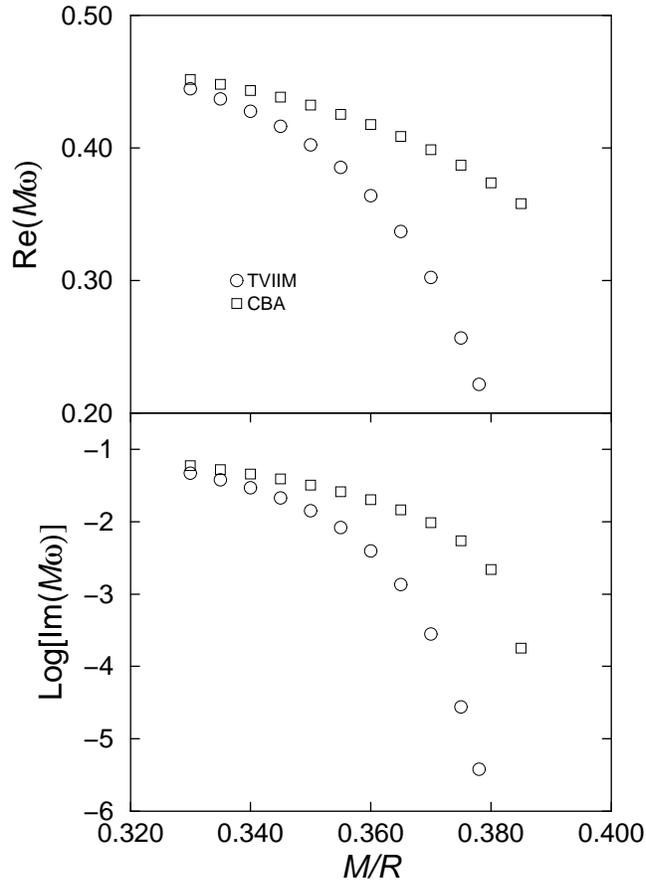}
 \caption{The real and imaginary parts of the scaled QNM
frequencies of a trapped-mode of TVIIM and CBA are plotted against
the compactness.} \label{f10}
\end{figure}

Despite that we have demonstrated here the validity of CBA and
SCLPT for ordinary $w$-modes and $w_{\rm II}$ modes; we have to
caution that the trapped mode is an exception to our method. As is
well known, trapped modes usually exist in ultra-compact stars
with compactness greater than 1/3 and have small decaying rates
\citep{trap1}. The physical origin of such modes is the
development of a local minimum in the potential $V(r_*)$ inside
the star when $\cc > 1/3$. Hence, gravitational waves are trapped
in the potential minimum and acquire much longer lifetime. The
analysis proposed in the present paper relies on the validity of
the CBA model whose potential term $V(r_*)$ is obviously a
monotonic function inside the star. In fact, the CBA potential
deviates significantly from those of TVIIM (or other ultra-compact
stars). As shown in Fig.~\ref{f10}, trapped modes of CBA are no
longer good approximation to those of TVIIM if $\cc > 1/3$ and
therefore it is not possible to obtain the trapped modes of TVIIM
(or other ultra-compact stars) from application of SCLPT to CBA.
We conclude our paper with this remark.

\begin{acknowledgments}
We thank J Wu for discussions. We also express our gratitude to
 an anonymous referee for drawing
our attention to the $w_{\rm II}$ and the trapped modes. Our work
is supported in part by the Hong Kong Research Grants Council
(grant No: CUHK4282/00P and 401905) and a direct grant (Project
ID: 2060260) from the Chinese University of Hong Kong.
\end{acknowledgments}

\newpage

\begin{thebibliography}{37}
\expandafter\ifx\csname
natexlab\endcsname\relax\def\natexlab#1{#1}\fi

\bibitem[{Akmal {et~al.}(1998)Akmal, Pandharipande, \& Ravenhall}]{APR}
Akmal, A., Pandharipande, V.~R., \& Ravenhall, D.~G. 1998, Phys.
Rev. C, 58,
  1804

\bibitem[{Alcock {et~al.}(1986)Alcock, Farhi, \& Olinto}]{AFO}
Alcock, C., Farhi, C.~E., \& Olinto, A. 1986, ApJ, 310, 261

\bibitem[{Andersson \& Kokkotas(1996)}]{Andersson_1996}
Andersson, N., \& Kokkotas, K.~D. 1996, Phys. Rev. Lett., 77, 20

\bibitem[{Andersson \& Kokkotas(1998)}]{Andersson1998}
---. 1998, MNRAS, 299, 1059

\bibitem[{Belczynski {et~al.}(2001)Belczynski, Kalogera, \&
  Bulik}]{Belczynski:2001uc}
Belczynski, K., Kalogera, V., \& Bulik, T. 2001, ApJ, 572, 407

\bibitem[{Benhar {et~al.}(1999)Benhar, Berti, \& Ferrari}]{Ferrari}
Benhar, O., Berti, E., \& Ferrari, V. 1999, MNRAS, 310, 797

\bibitem[{Chandrasekhar \& Ferrari(1991a)}]{Chandrasekhar1}
Chandrasekhar, S., \& Ferrari, V. 1991a, Proc. R. Soc. A, 432, 247

\bibitem[{Chandrasekhar \& Ferrari(1991b)}]{trap1}
---. 1991b, Proc. R. Soc. A, 434, 449

\bibitem[{Cheng {et~al.}(1998)Cheng, Dai, Wei, \& Lu}]{Cheng}
Cheng, K.~S., Dai, Z.~G., Wei, D.~M., \& Lu, T. 1998, Science,
280, 407

\bibitem[{Ching {et~al.}(1996)Ching, Leung, Suen, \& Young}]{Ching}
Ching, E.~S.~C., Leung, P.~T., Suen, W.~M., \& Young, K. 1996,
Phys. Rev. D,
  54, 3778

\bibitem[{Chodos {et~al.}(1974)Chodos, Jaffe, Johnson, Thorne, \&
  Weisskopf}]{MITBM}
Chodos, A., Jaffe, R.~L., Johnson, K., Thorne, C.~B., \&
Weisskopf, V.~F. 1974,
  Phys. Rev. D, 9, 3471

\bibitem[{Fryer {et~al.}(2002)Fryer, Holz, \& Hughes}]{Fryer:2001zw}
Fryer, C.~L., Holz, D.~E., \& Hughes, S.~A. 2002, ApJ, 565, 430

\bibitem[{Glendenning(1997)}]{ComStar}
Glendenning, N.~K. 1997, Compact Stars - Nuclear Physics, Particle
Physics, and
  General Relativity (Springer, NY)

\bibitem[{Hughes(2003)}]{Hughes_03}
Hughes, S. 2003, Ann. Phys., 303, 142

\bibitem[{Kokkotas {et~al.}(2001)Kokkotas, Apostolatos, \&
  Andersson}]{Kokkotas_2001}
Kokkotas, K.~D., Apostolatos, T.~A., \& Andersson, N. 2001, MNRAS,
320, 307

\bibitem[{Kokkotas \& Schmidt(1999)}]{Kokkotas_rev}
Kokkotas, K.~D., \& Schmidt, B.~G. 1999, Living Rev. Rel., 2, 2

\bibitem[{Kokkotas \& Schutz(1986)}]{ToyModel}
Kokkotas, K.~D., \& Schutz, B.~F. 1986, Gen. Relativ. Gravitation,
18, 913

\bibitem[{Lattimer \& Prakash(2001)}]{Lattimer:2001}
Lattimer, J.~M., \& Prakash, M. 2001, ApJ, 550, 426

\bibitem[{Leaver(1986{\natexlab{a}})}]{Leaver_1986}
Leaver, E.~W. 1986{\natexlab{a}}, Phys. Rev. D, 34, 384

\bibitem[{Leaver(1986{\natexlab{b}})}]{Leaver_series}
---. 1986{\natexlab{b}}, J. Math. Phys., 27, 1238

\bibitem[{Leung {et~al.}(1997)Leung, Liu, Suen, Tam, \& Young}]{dirty1}
Leung, P.~T., Liu, Y.~T., Suen, W.~M., Tam, C.~Y., \& Young, K.
1997, Phys.
  Rev. Lett., 78, 2894


\bibitem[{Leung {et~al.}(1999)Leung, Liu, Suen, Tam, \& Young}]{dirty2}
---. 1999, Phys. Rev. D, 59, 044034

\bibitem[{Lindblom(1992)}]{Lindblom_invert}
Lindblom, L. 1992, ApJ, 398, 56

\bibitem[{Lindblom {et~al.}(1998)Lindblom, Owen, \& Morsink}]{Lindblom:1998wf}
Lindblom, L., Owen, B.~J., \& Morsink, S.~M. 1998, Phys. Rev.
Lett., 80, 4843

\bibitem[{Liu(1997)}]{Liu}
Liu, Y.~T. 1997, MPhil thesis, The Chinese University of Hong Kong

\bibitem[{Mason(2004)}]{Grishchuk}
Mason, J.~W., ed. 2004, Astrophysics Update, Vol.~I
(Springer-Praxis), 281--310

\bibitem[{Nollert(1999)}]{Nollert_rev}
Nollert, H.-P. 1999, Class. Quantum Grav., 16, R159

\bibitem[{Pandharipande(1971{\natexlab{a}})}]{modelA}
Pandharipande, V. 1971{\natexlab{a}}, Nucl. Phys A, 174, 641

\bibitem[{Pandharipande(1971{\natexlab{b}})}]{modelC}
---. 1971{\natexlab{b}}, Nucl. Phys A, 178, 123

\bibitem[{Prakash {et~al.}(1990)Prakash, Baron, \& Prakash}]{PBP}
Prakash, M., Baron, E., \& Prakash, M. 1990, Phys. Lett. B, 243,
175

\bibitem[{Press(1971)}]{Press_1971}
Press, W.~H. 1971, ApJ, 170, L105

\bibitem[{Regge \& Wheeler(1957)}]{RWeq}
Regge, T., \& Wheeler, J.~A. 1957, Phys. Rev., 108, 1063

\bibitem[{Thorne \& Campolattaro(1967)}]{Thorne}
Thorne, K.~S., \& Campolattaro, A. 1967, ApJ, 149, 591

\bibitem[{Tolman(1939)}]{Tolman:1939jz}
Tolman, R.~C. 1939, Phys. Rev., 55, 364

\bibitem[{Tsui \& Leung(2004)}]{preprint}
Tsui, L.~K., \& Leung, P.~T. 2004, MNRAS, 357, 1029

\bibitem[{Wiringa {et~al.}(1988)Wiringa, Fiks, \& Fabrocini}]{AU}
Wiringa, R.~B., Fiks, V., \& Fabrocini, A. 1988, Phys. Rev. C, 38,
1010

\bibitem[{Witten(1984)}]{Witten}
Witten, E. 1984, Phys. Rev. D, 30, 272

\end{thebibliography}
\newcommand{\noopsort}[1]{} \newcommand{\printfirst}[2]{#1}
  \newcommand{\singleletter}[1]{#1} \newcommand{\switchargs}[2]{#2#1}

\newpage



\clearpage 
\begin{table}
  \centering
  \begin{tabular}{|c|c|c|c|}
    \hline
    Data & $a$ & $b$ & $c$\\
    \hline
    Best fit to realistic stars & $-3.9-\i5.6$ & $2.8+\i1.6$ & $-0.03+\i0.125$ \\
    Perturbative result of CBA& $-4.63-\i4.86$ & $3.15+\i1.30$ & $-0.063+\i0.155$ \\
    \hline
  \end{tabular}
  \caption{The complex coefficients $a$, $b$ and $c$ in (\ref{TLA}) are shown.
  In the first row the values are obtained from the best quadratic fit to
   the QNMs of the realistic stars considered in Fig.~\ref{f1},
   while in the second row those obtained from the second
  order result of SCLPT for
  a reference CBA star with compactness $M/R=0.2$ are presented.
  }
  \label{coef_cea_cba}
\end{table}
\clearpage
\end{document}